\documentclass[
floatfix,
aps,
prl,
twocolumn,
superscriptaddress,
amssymb,
 groupaddress,
]{revtex4}

\usepackage{float}
\usepackage{amssymb}
\usepackage{gensymb}
\usepackage{latexsym}
\usepackage[pdftex]{graphicx}
\usepackage{amsmath}
\usepackage{graphicx}
\usepackage{dcolumn}
\usepackage{amsfonts}
\usepackage{bm}
\usepackage{epsfig}

\def\nn{\nonumber}

\def\nt{\nu_{\mathrm{T}}}

\def\nt{n_{\mathrm{T}}}

\def\ar{\alpha_{\mathrm{r}}}

\def\np{n_{+}}
\def\nn{n_{-}}
\def\nt{n_{\mathrm{T}}}
\def\lso{\ell_{\mathrm{SO}}}

\begin{document}
\title{Mobility in excess of $10^{6}$ cm$^2$/Vs in InAs quantum wells grown on lattice mismatched InP substrates}

\author{A.\,T. Hatke}
\affiliation{Department of Physics and Astronomy and Station Q Purdue, Purdue University, West Lafayette, Indiana 47907, USA}

\author{T. Wang}
\affiliation{Department of Physics and Astronomy and Station Q Purdue, Purdue University, West Lafayette, Indiana 47907, USA}
\affiliation{Birck Nanotechnology Center, Purdue University, West Lafayette, IN, 47907 USA}

\author{C. Thomas}
\affiliation{Department of Physics and Astronomy and Station Q Purdue, Purdue University, West Lafayette, Indiana 47907, USA}
\affiliation{Birck Nanotechnology Center, Purdue University, West Lafayette, IN, 47907 USA}

\author{G. Gardner}
\affiliation{Department of Physics and Astronomy and Station Q Purdue, Purdue University, West Lafayette, Indiana 47907, USA}
\affiliation{Birck Nanotechnology Center, Purdue University, West Lafayette, IN, 47907 USA}
\affiliation{School of Materials Engineering, Purdue University, West Lafayette, IN, 47907 USA}

\author{M. J. Manfra}
\affiliation{Department of Physics and Astronomy and Station Q Purdue, Purdue University, West Lafayette, Indiana 47907, USA}
\affiliation{Birck Nanotechnology Center, Purdue University, West Lafayette, IN, 47907 USA}
\affiliation{School of Materials Engineering, Purdue University, West Lafayette, IN, 47907 USA}
\affiliation{School of Electrical and Computer Engineering School of Materials Engineering Purdue University, West Lafayette, IN 47907, USA}
 
\received{\today}

\begin{abstract}
InAs-based two-dimensional electron systems grown on lattice mismatched InP substrates offer a robust platform for the pursuit of topologically protected quantum computing.  We investigated strained composite quantum wells of In$_{0.75}$Ga$_{0.25}$As/InAs/In$_{0.75}$Ga$_{0.25}$As with In$_{0.75}$Al$_{0.25}$As barriers. 
By optimizing the widths of the In$_{0.75}$Ga$_{0.25}$As layers, the In$_{0.75}$Al$_{0.25}$As barrier, and the InAs quantum well we demonstrate mobility in excess of $1 \times 10^{6}\,$cm$^{2}/$Vs.  Mobility vs. density data indicates that scattering is dominated by a residual three dimensional distribution of charged impurities. We extract the Rashba parameter and spin-orbit length as important material parameters for investigations involving Majorana zero modes.
\end{abstract} 

\maketitle

Due to a combination of strong spin-orbit coupling, large g-factor, and readily-induced proximity superconductivity the InAs based two-dimensional electron gas (2DEG) has gained traction recently as a promising platform for topological quantum computing \citep{sarma:2005,alicea:2011,shabani:2016,kjaergaard:2016,kjaergaard:2017,suominen:2017,nichele:2017}.
A structure composed of a shallow InAs quantum well can be engineered to have proximity induced superconductivity with an in-situ epitaxial Al top layer with high transparency \citep{shabani:2016}.
This system has been demonstrated to contain Andreev bound states that coalesce into Majorana zero modes \citep{suominen:2017,nichele:2017}.
However, a pressing limitation is the quality of the 2DEG.

We investigate the limitations of 2DEG mobility in the InAs on InP substrate system.
Low temperature transport measurements are performed on gated Hall bars using symmetric In$_{0.75}$Ga$_{0.25}$As/InAs/In$_{0.75}$Ga$_{0.25}$As quantum wells grown on (100) InP where we vary the width of the flanking InGaAs layers, the depth of the quantum well from the surface, and the width of the InAs layer. 
While InAs has a 3.3$\%$ lattice mismatch to InP, the superior insulating property of Fe-doped InP substrates presents a crucial advantage for the measurement of high impedance devices necessary for the exploration of Majorana physics.
Our results demonstrate record charge carrier mobility in excess of $1 \times 10^{6}\,$cm$^{2}/$Vs for this system and that our mobility appears to be limited by unintentional background charge impurities.
We extract the Rashba parameter and spin-orbit length from the beating pattern of Shubnikov de-Haas oscillations.
These results may be leveraged to improve the quality of InAs 2DEG structures used for topological quantum computing.

\begin{table}[b]
  \begin{tabular}{ | l | c | c | c | }
    \hline
    \, & \,\,In$_{0.75}$Al$_{0.25}$As\,\, & \,\,In$_{0.75}$Ga$_{0.25}$As\,\, & \,\,InAs\,\, \\ \hline
    \,\,Sample A\,\, & $b=120\,$nm & $d=5\,$nm & $w=4\,$nm \\ \hline
    \,\,Sample B\,\, & $b=120\,$nm & $d=10.5\,$nm & $w=4\,$nm \\ \hline
    \,\,Sample C\,\, & $b=120\,$nm & $d=15\,$nm & $w=4\,$nm  \\ \hline
    \,\,Sample D\,\, & $b=120\,$nm & $d=10.5\,$nm & $w=6\,$nm \\ \hline
    \,\,Sample E\,\, & $b=180\,$nm & $d=10.5\,$nm & $w=4\,$nm \\ 
    \hline
  \end{tabular}
  \caption{Sample details for dimensions of the top In$_{0.75}$Al$_{0.25}$As barrier layer,  the symmetric In$_{0.75}$Ga$_{0.25}$As layers, and the InAs quantum well width.}
\label{table}
\end{table}

Our samples are grown using molecular beam epitaxy (MBE); see Ref.\,\citep{gardner:2016} for greater detail on how our MBE has been set up and maintained.
The InAs structures are grown on semi-insulating InP (100) substrates that have been desorbed at $525^{\degree}$C until a $(2\times3)$ to $(2\times4)$ surface phase transition is observed.
First, a 100 nm thick In$_{0.52}$Al$_{0.48}$As lattice matched smoothing layer upon which a In$_{0.52}$Al$_{0.48}$As/In$_{0.52}$Ga$_{0.48}$As  $2.5\,$nm five period superlattice is grown at $480^{\degree}$C \citep{heyn:2003}.
Due to a native lattice mismatch of 3.3$\%$ between InAs and InP we grow a step graded buffer of In$_{x}$Al$_{1-x}$As where $x=0.52$ to $0.84$ using 18, $50\,$nm wide each, followed by a linearly ramped reverse step from $x=0.84$ to $0.75$ to relieve any residual strain.  
The graded buffer layer and reverse step are grown at $360^{\degree}$C.

\begin{figure}[t]
\vspace{-0.1 in}
\includegraphics[width=.48\textwidth]{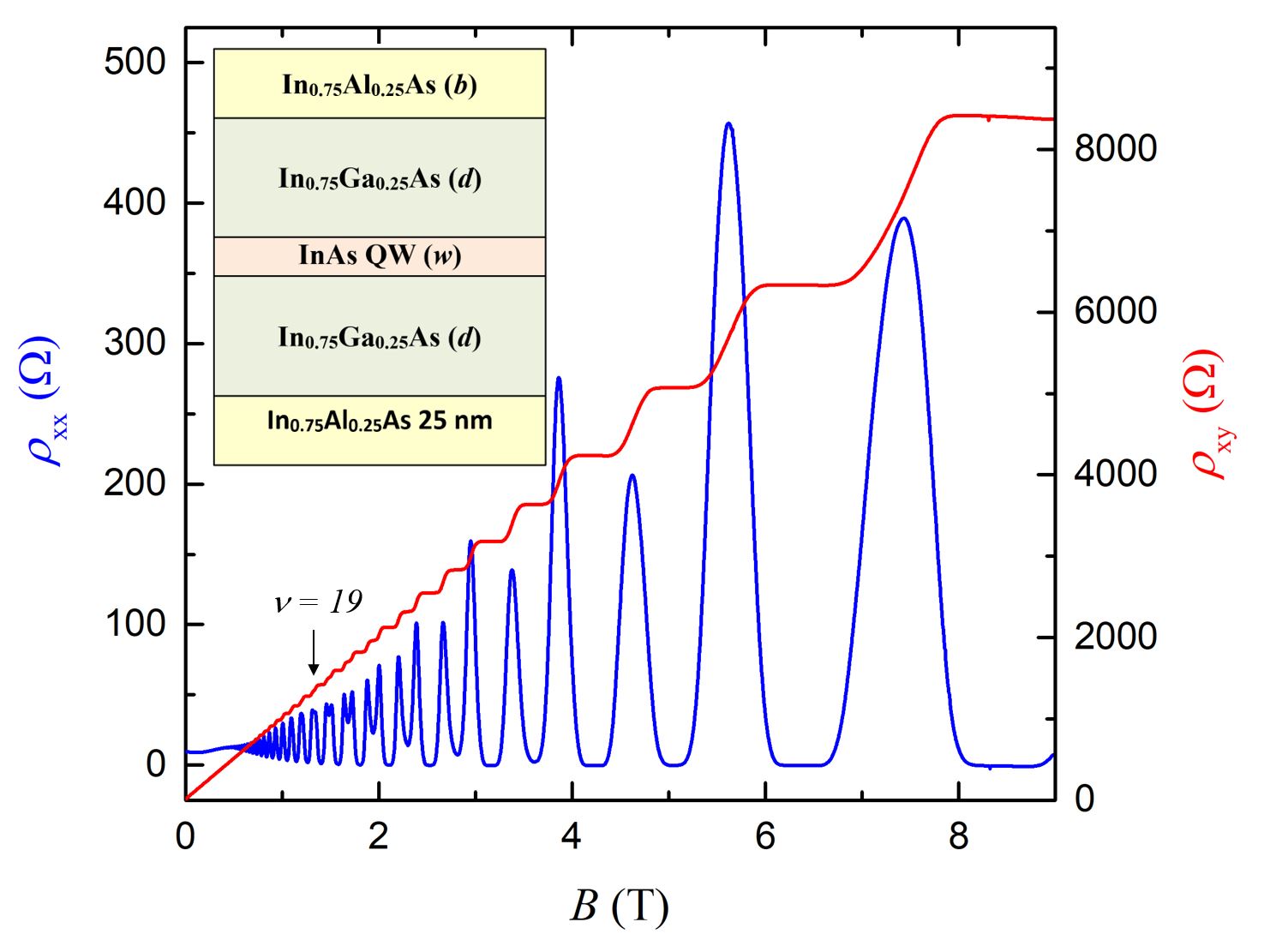}
\vspace{-0.25 in}
\caption{(Color online)
Longitudinal, $\rho_{xx}$, and Hall, $\rho_{xy}$ in units of inverse filling factor, resistivities vs magnetic field, $B$, for density $n=6.2\times 10^{11}\,$cm$^{-2}$, left and right axis respectively, for Sample B.
Inset: Schematic representation of the layer stack for the the active region of the quantum well, see text for greater detail.
}
\vspace{-0.1 in}
\label{fig1}
\end{figure}

The active region comprised of the composite quantum well plus barriers is then grown. The substrate temperature is increased to $480^{\degree}$C to grow a $25\,$nm In$_{0.75}$Al$_{0.25}$As bottom barrier and active region composed of a strained $w=4\,$nm ($w=6\,$nm for Sample E only) InAs layer flanked on either side by symmetric In$_{0.75}$Ga$_{0.25}$As layers to promote higher mobility \citep{sexl:1997,wallart:2005}.
For Samples A, B, and C, we vary only the width of the In$_{0.75}$Ga$_{0.25}$As layers to be $d=5, 10.5,$ and $15\,$nm, respectively.
The sample growth is completed with a $b=120\,$nm ($b=180\,$nm for Sample D only) In$_{0.75}$Al$_{0.25}$As top barrier to remove the active region from the surface and minimize anisotropy effects that can become apparent when the active region of the quantum well is to near the surface \citep{lohr:2003}.
Lastly, we do not include an InGaAs capping layer to avoid formation of a parallel conduction channel \citep{shabani:2014} or intentional doping.
In the inset of Fig.\,\ref{fig1} we schematically depict the layer stack for the active region.  
A summary of the five samples discussed here are presented in Table\,\ref{table}.

Our samples are processed with standard wet etching techniques to define both straight and L-shaped (aligned along the $[1\bar{1}0]$ and $[110]$ directions) Hall bars of width $w=150\,\mu$m. 
After etching we deposit Ti/Au ohmic contacts of thickness $80/250\,$nm, a 40 nm Al$_{2}$O$_{3}$ dielectric using thermal atomic layer deposition, and a $20/150\,$nm Ti/Au gate.
All samples have a zero gate voltage, $V_{\mathrm{G}}=0$, density of  $n = 5.3-5.6 \times 10^{11}\,$cm$^{-2}$ with $\Delta n$ versus $V_{\mathrm{G}}$ in good agreement with a simple capacitance model. 
The samples were measured in a $^{3}$He system at a base temperature of $T=300\,$mK using standard low frequency lockin techniques with excitation current of $0.5 \mu$A.

InAs quantum wells based on GaSb substrates with Al$_{0.37}$Ga$_{0.67}$Sb barriers \citep{shojaei:2016} have recently been shown to achieve mobilities of $\mu=2.4 \times 10^{6}\,$cm$^{2}/$Vs  at $n \sim 1 \times 10^{12}\,$cm$^{-2} $\citep{tschirky:2017}.
The sample structures investigated here are instead grown on lattice mismatched InP substrates that have superior insulating properties, a requirement when operating mesoscopic devices in high resistance configurations.
To our knowledge, the highest reported mobility for such a structure is $\mu= 0.6 \times 10^{6}\,$cm$^{2}/$Vs achieved at $n \sim 5 \times 10^{11}\,$cm$^{-2}$ \citep{shabani:2014}, but supporting transport data was not provided.

We begin our discussion with Sample B, which yielded the highest mobility.
In Fig.\,\ref{fig1} we present longitudinal ($\rho_{xx}$, left axis) and Hall ($\rho_{xy}$ in units of inverse filling factor, right axis) resistivities versus magnetic field ($B$) for $n = 6.2 \times 10^{11}\,$cm$^{-2}$.
We observe the absence of a parasitic parallel conduction channel from the linear low field $\rho_{xy}$.
There is also good agreement between the extracted density from both the Hall slope and the period of Shubnikov de-Haas oscillations (SdHOs). 
With increasing $B$ we observe a spin splitting onset at filling factor $\nu=nh/eB=19$, where $h$ is the Planck constant and $e$ the electron charge, as marked in Fig.\,\ref{fig1}, with well developed integer quantum Hall states, $\rho_{xx} = 0$ and $\rho_{xy} =N h/e^{2}\nu$, where $N$ is an integer.

\begin{figure}[t]
\vspace{-0.1 in}
\includegraphics[width=.48\textwidth]{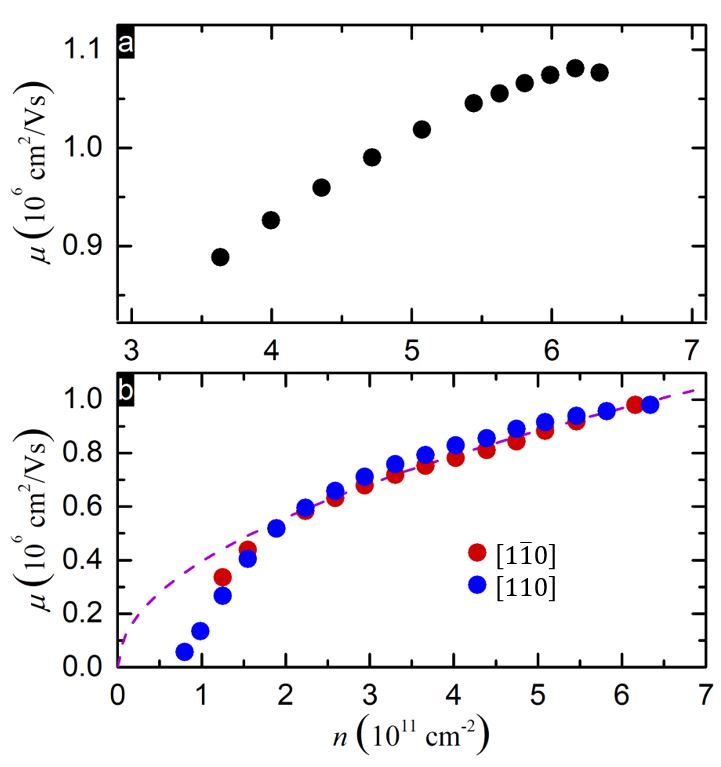}
\vspace{-0.25 in}
\caption{(Color online)
Mobility vs density, $\mu$ vs $n$:
(a) Sample B on a straight Hall bar oriented along the $[1\bar{1}0]$ direction demonstrating a peak mobility of $\mu=1.1 \times 10^{6}\,$cm$^{2}/$Vs at $n = 6.2 \times 10^{11}\,$cm$^{-2}$.
(b) Sample B obtained from an L-shaped Hall bar along the two main crystallographic directions and a fit line to $\mu \propto n^{\alpha}$ where $\alpha=0.5$.
}
\vspace{-0.1 in}
\label{fig2}
\end{figure}
We continue with Sample B with gating to obtain $\mu$ versus $n$, shown in Fig.\,\ref{fig2}\,(a) for a straight Hall bar aligned along the $[1\bar{1}0]$ direction.
A maximum mobility of $\mu=1.1 \times 10^{6}\,$cm$^{2}/$Vs occurs at $n = 6.2 \times 10^{11}\,$cm$^{-2}$.
To our knowledge this is the largest reported $\mu$ for an InGaAs/InAs/InGaAs quantum well. 
On a device processed during a different fabrication, from the same wafer as Sample B, we plot $\mu$ versus $n$ where measurements were performed on an L-shaped Hall bar in Fig.\,\ref{fig2}\,(b).
The gating dependence shows that there is minimal anisotropy between the $[1\bar{1}0]$ and $[110]$ directions with less than $5\%$ difference. 
Thus we compare samples only along the $[1\bar{1}0]$ direction for the remainder of the manuscript.

To determine what limits mobility in our structure we assume the $\mu$ vs $n$ dependence can be described by a simple power law, $\mu\propto n^{\alpha}$, and extract the exponent $\alpha$, using a log-log plot (not shown) giving equal weight to all data, in the restricted density range of $n>1.5 \times 10^{11}\,$cm$^{-2}$.
In Fig.\,\ref{fig2}\,(b) the extracted fit, dashed line, for $\alpha = 0.5$ fits well over the density range of interest and is roughly equivalent for all samples measured in this study.
In Ref.\,\citep{shabani:2014_MIT} a similar structure was investigated that contained $8\,$nm In$_{0.75}$Ga$_{0.25}$As layers where it is was observed that $\alpha \sim 0.8$.
These $\alpha$-values indicate that the mobility is limited by unintentional background impurities \citep{sarma:2013}.
Theoretically, in the strong screening, $q_{\mathrm{TF}} \gg k_{\mathrm{F}}$ where $q_{\mathrm{TF}}$ is the Thomas-Fermi wave vector \citep{stern:1967}, and high density limit $\alpha \rightarrow 1/2$, however, in the case of remote two-dimensional impurities $\alpha \rightarrow 3/2$.
For comparison, Ref.\,\citep{shabani:2014_MIT} investigated a sample with an additional $10\,$nm In$_{0.75}$Ga$_{0.25}$As capping layer and observed $\alpha = 1.35$, where the increase in $\alpha$ was attributed to an unintentional parallel surface channel. 
It is plausible that the introduction of this capping layer enhanced a remote layer that was competing with the background impurities favoring an increase in $\alpha$.
The difference between $\alpha=0.5$ and $\alpha=0.8$ could also be due to unintentional background impurities within the well, as evidenced by the difference in $\mu$ over the same $n$ range \citep{sarma:2013}.
At present the exact nature of the charged impurities in our samples cannot be specified, but since a 2DEG is formed in the absence of modulation doping it is reasonable to assume that ionized donor-like defects exist in the lattice.  
Identification of the precise location and density of such defects requires further investigation beyond the scope of this paper.

\begin{figure}[t]
\vspace{-0.1 in}
\includegraphics[width=.48\textwidth]{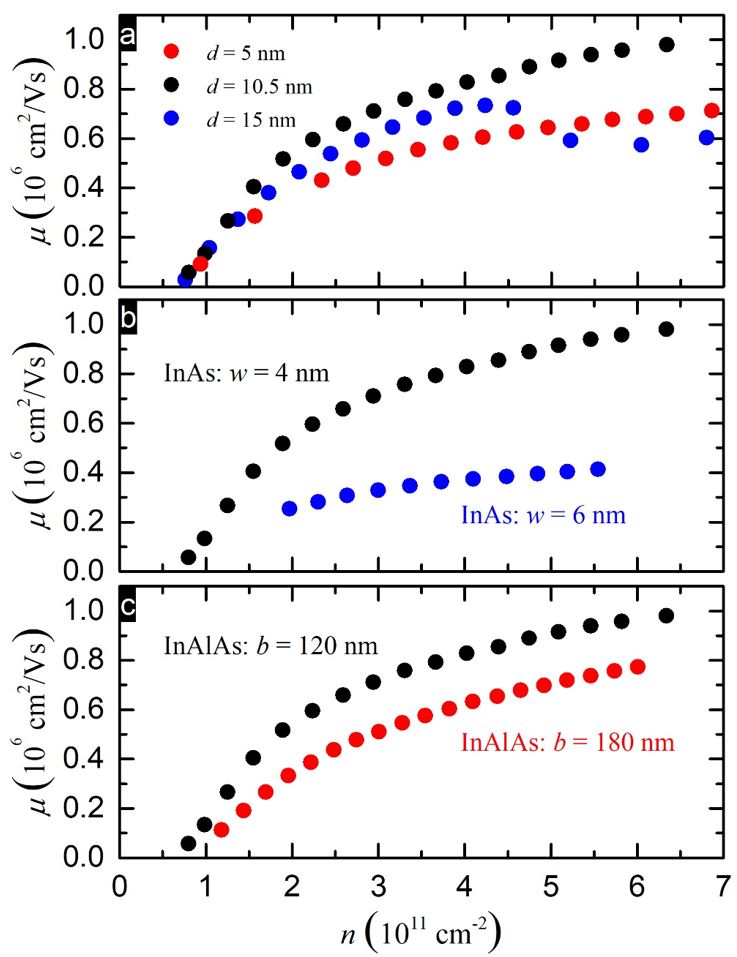}
\vspace{-0.25 in}
\caption{(Color online)
$\mu$ vs $n$:
(a) Comparison from Samples A-C where the width, $d$, of the InGaAs layer is varied.
(b) Comparison between Sample B and D where the width, $w$, of the InAs quantum well is varied.
(c) Comparison between Sample B and E where the width, $b$, of the top InAlAs barrier is varied.
} 
\vspace{-0.1 in}
\label{fig3}
\end{figure}
Comparison of the quality of our samples is evaluated using zero-field mobility as the metric.
We next investigate perturbations to Sample B beginning with the well width dependence of the In$_{0.75}$Ga$_{0.25}$As layers.
In Fig.\,\ref{fig2}\,(c) we plot $\mu$ versus $n$ for Samples A-C where the In$_{0.75}$Ga$_{0.25}$As layer widths are $d=5,10.5,\,$and $15\,$nm, respectively.
For Sample C, $d=15\,$nm, there is a nonmonotonic $\mu$ versus $n$ where $\mu$ begins to decrease for $n > 4.25 \times 10^{11}\,$cm$^{-2}$.
This nonmonotonic $n$-dependence is due to occupation of the second subband.
Estimation of the onset density of the second subband becoming populated occurs at $n \sim 7.5, 6,$ and $5 \times 10^{11}\,$cm$^{-2}$ for Samples A-C, respectively, from self consistent calculations performed with Nextnano$^{3}$ \citep{nextnano}.

At fixed $n$ there is a nonmonotonic dependence of $\mu$ versus $d$.
At $n\sim 4 \times 10^{11}\,$cm$^{-2}$, for example, $\mu= 0.83 \times 10^{6}\,$cm$^{2}/$Vs for $d=10.5\,$nm that decreases to $\mu=0.73 \times 10^{6}\,$cm$^{2}/$Vs and $\mu=0.59 \times 10^{6}\,$cm$^{2}/$Vs for $d=15$ and $5\,$nm, respectively.
With a large overlap in sample structure between the three samples we do not expect changes in scattering from background impurities, remote impurities, or charged dislocations due to the lattice mismatch to give reasonable explanation to the observed width dependence.
Increasing $d$ results in a spread of the charge distribution such that there is an increase of the amount of charge that resides in the In$_{0.75}$Ga$_{0.25}$As layers.
An increase of the amount of wavefunction extension into the In$_{0.75}$Ga$_{0.25}$As layer will decrease the mobility due to an increase in the amount of alloy scattering.
The small decrease of $\mu$ of $\sim 12\%$ is due to a $\sim 1\%$ transfer of charge from the pure InAs to the In$_{0.75}$Ga$_{0.25}$As layer implies a strong dependence on alloy scattering.
A more dramatic reduction in the mobility occurs when there is a decrease of $d$, which can come from two sources 1) alloy scattering and 2) interface scattering.
The charge distribution in the effective $14\,$nm well of Sample C will penetrate into the In$_{0.75}$Al$_{0.25}$As barriers giving an increased amount of alloy scattering, as observed in Nextnano$^{3}$ simulations.
Additionally, the increased confinement of the charge results in an increase in scattering at the In$_{0.75}$Ga$_{0.25}$As/InAs interface.
Comparison of the integrated charge density of the wells of Sample B and C in a restricted region of $0.5\,$nm to either side of the InGaAs/InAs interface shows that the amount of charge in the region of the interface increases giving rise to increased interface scattering.

In Fig.\,\ref{fig3}\,(b) $\mu$ versus $n$ for Sample B and D where the width of the InAs quantum well is increased from $w=4$ to $6\,$nm is plotted.
A large reduction in $\mu$ throughout the entire $n$-range for $w=6\,$nm is observed.
Naively one might expect $\mu$ to increase with an increase in $w$ as a larger percentage of the charge density would reside in the InAs part of the well resulting in a decrease in alloy scattering from the In$_{0.75}$Ga$_{0.25}$As layers.
However, the severe reduction in $\mu$ implies that the critical thickness, $w_{\mathrm{c}}$, of the InAs has been exceeded which introduce misfit dislocations to the quantum well \citep{capotondi:2005,shabani:2014}.
We estimate $w_{\mathrm{c}}\sim 5.5\,$nm, for this In concentration.

In Fig.\,\ref{fig3}\,(c) we plot $\mu$ versus $n$ for increase of the In$_{0.75}$Al$_{0.25}$As barrier from $b=120$ to $180\,$nm, Samples B and E respectively.
Again there is an overall decrease in $\mu$.
As previously discussed $\mu$ is limited by background charged impurities which suggests that while $b$ is increased in Sample E to reduce surface effects the possible gain is compensated by the increased level of charged impurities introduced by the additional In$_{0.75}$Al$_{0.25}$As layers resulting in decreased $\mu$.

\begin{figure}[t]
\vspace{-0.1 in}
\includegraphics[width=.5\textwidth]{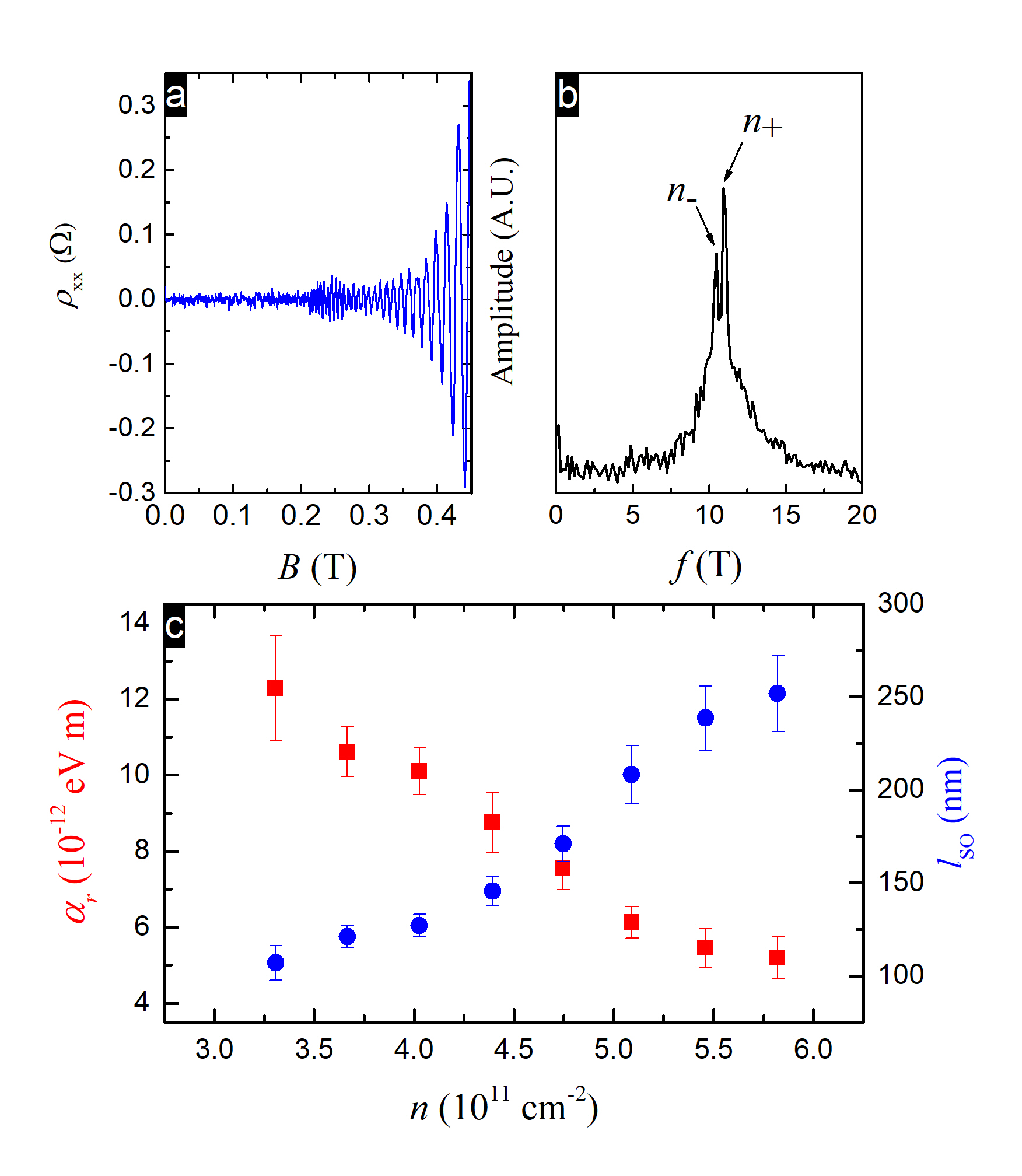}
\vspace{-0.25 in}
\caption{(Color online)
(a) Low field magnetoresistivity with removal of a smoothly varying background, $\Delta \rho_{xx}$ vs $B$, for $n=5.1 \times 10^{11}\,$cm$^{2}$ on Sample B.
(c) Amplitude of a FFT of $\rho_{xx}$ from inverse $B$.
(c) Rashba parameter, $\ar$ squares, and spin orbit length, $\lso$ circles, vs $n$, left and right axis respectively.
}
\vspace{-0.1 in}
\label{fig4}
\end{figure}
We preformed further measurements of Sample B at low $B$ to investigate the spin-orbit coupling.
In Fig.\,\ref{fig4}\,(a) we plot the oscillatory correction to the magnetoresistivity, $\Delta\rho_{\mathrm{xx}}$, for $n=5.1 \times 10^{11}\,$cm$^{-2}$ after removal of a slowly varying background.
With increasing $B$ we observe the onset of SdHOs at $B\sim 0.2\,$T, this low $B$ value is another indication of the high quality of the 2DEG.
The amplitude of the SdHOs increase with increasing $B$ but demonstrate a beating pattern with a node at $B \sim 0.3\,$T.
By restricting our analysis to densities below occupation of the second subband, this beating pattern can be ascribed to two oscillation periods that are nearly equal and has been demonstrated in these structures to arise from zero field spin splitting between slightly different spin up and spin down densities \citep{datta:1990,kim:2010,lee:2011}.

In Fig.\,\ref{fig4}\,(b) we present the amplitude of the Fast Fourier Transform (FFT) versus frequency of the magnetotransport after conversion to inverse magnetic field.
This FFT split peak can be assigned to two spin-split subbands with densities $\np$ and $\nn$, which can be calculated from $n_{\pm}=ef/h$.
From this assignment, the estimated total density $\nt=\np+\nn$ is in good agreement with that obtained from the Hall slope and the SdHO minima period.

In systems that lack inversion symmetry the dominant source of spin-orbit interaction is due to the Rashba effect, which arises from an electric field perpendicular to the plane of the 2DEG.
This electric field can be a result of an inversion asymmetry built into the system based on 2DEG design or from an applied field from a gate \citep{nitta:1997}.
From the SdHO beating pattern we extract the Rashba parameter $\ar=\frac{\Delta n\hbar^{2}}{m^{*}}\sqrt{\frac{\pi}{2(\nt-\Delta n)}}$, where $\Delta n = \np-\nn$ and we assume $m^{*}=0.03$ \citep{shabani:2014_MIT}.
We perform FFTs at different $V_{\mathrm{G}}$ and extract $\ar$ versus $n$ in Fig.\,\ref{fig4}\,(c), left axis.
The Rashba effect is due to an asymmetry in the azimuthal direction and is proportional to the electric field, $\ar = \alpha_{\mathrm{0}}\langle E_{\mathrm{z}}\rangle$ where $\alpha_{\mathrm{0}}$ is a material specific parameter.
Our gating density dependence is very nearly linear and follows from the simple capacitance model where we observe a linear change to $n$ so we expect a linear increase of $\ar$ with decreasing $n$, corresponding to an increase in $E_{\mathrm{Z}}$.
The values we obtain for $\ar$ are of the same order as those obtained from InAs systems with symmetric Si doping \citep{kim:2010}, built in In$_{0.53}$Ga$_{0.47}$As layer asymmetry \citep{lee:2011,park:2013}, or those reported with AlSb barriers \citep{shojaei:2016}.

To eliminate effective mass dependence we recast $\ar$ as the spin-orbit length, $\lso=\frac{1}{\Delta n}\sqrt{\frac{\nt-\Delta n}{2\pi}}$, versus $n$ and plot the result in Fig.\,\ref{fig4}\,(c), right axis.
Physically, the spin-orbit length gives a measure of the average distance traversed by an electron before a spin flip occurs.
In the case of weak spin-orbit interaction, the high $n$ (low $V_{\mathrm{G}}$) case, the spin will travel further through the system, larger $\lso$, before its spin orientation will become essentially randomized.
With increase of the spin-orbit interaction under applied gate voltage the electron traverses decreasing distance before its spin is randomized.

This research supported by Microsoft Station Q.

\bibliographystyle{apsrev}

\end{document}